\documentclass[aps,pre,twocolumn,showpacs]{revtex4}
\usepackage{amsmath,amsthm}  
\usepackage{epsf}
\usepackage{amssymb}  
\usepackage{graphicx}

\newcommand{\eq}[1]{Eq.~(\ref{#1})}
\newcommand{\be}{\begin{eqnarray}}
\newcommand{\ee}{\end{eqnarray}}
\newcommand{\vecn}{\mathbf{n}}
\newcommand{\veck}{\mathbf{k}}

\newcommand{\vece}{\mathbf{e}}

\newcommand{\vecgamma}{{\boldsymbol \gamma}}
\newcommand{\vecR}{\mathbf{R}}

\newcommand{\vecx}{\mathbf{x}}

\begin{document}
\title{Feedback between interacting transport channels}
\author{T. Brandes}
\affiliation{  Institut f\"ur Theoretische Physik,
  Hardenbergstr. 36,
  TU Berlin,  D-10623 Berlin,
  Germany}
\date{\today{ }}
\begin{abstract}
A model for autonomous feedback control of  particle transport through a large number of channels is introduced. Interactions among the particles
can lead to a strong suppression of fluctuations in the particle number statistics. Within a mean-field type limit, the collective control mechanism 
becomes equivalent to a  synchronization with an external clock. The diffusive spreading of the feedback signal across the channels shows scaling, can be quantified via the flow of information,  and 
shows up, e.g., in the spectral function of the particle noise.
\end{abstract}
\maketitle

\section{Introduction}
Feedback loops are an interesting tool to modify and control the transport dynamics of systems, both classical and quantum \cite{Zhangetal2015,WM2009,TSUMS2010,SEKB11,AMP2011,AS2011,JB2012,BGK2013,Kosetal2014,LGK2014}. Increasing efforts have been made recently to understand and quantify closed loop control schemes (e.g. in Maxwell demon, information tape or network  models) from the perspective of thermodynamics, statistical mechanics and system-bath theories \cite{SU2008,SU2010,HP2011,SU2012,AS2012,MR2012,ES2012,HSP2013,SSEB2013,Tas2013,HE2014,HBS2014,BS2014,SS2015,SSBJ2014,Hor2015,Gri2015}. 

A key question here is how to design the feedback loop itself, i.e. whether to regard it as part of some active external measurement scheme, or `passively' as an extended  part of the system itself. This latter form  of coherent control \cite{Lloyd2000,Caretal2013,HSCK2014,GPS2014,KESB2015,EG2014} is particularly appealing for quantum systems (also cf. \cite{EG2014} for coherent feedback control of quantum transport and further references), as it avoids the need to involve the measurement process. But also classically, a full microscopic understanding 
of feedback control requires the modeling of some form of interaction between the system and its controller. 

It is this perspective from which we re-visit a feedback control scheme \cite{Bra10} of a stochastic process describing (source to drain) transport of particles, which in its simplest version describes 
a continuous, active modulation of transition rates conditioned upon the full counting statistics (FCS) \cite{Nazarov2013}, i.e. the statistics $p(n,t)$ of particle numbers $n$ that have accumulated in a drain reservoir after a certain time $t$. In the new feedback model, this modulation is achieved in passive mode via interactions among particles in $N$ coupled transport `channels' that provide the  feedback by making  the individual transition rates in each channel dependent upon the  (reservoir) state of the entire system. 

The starting point in this paper thus is an infinite system of coupled Poissonian processes  for the FCS obtained  by adiabatic elimination of internal `connector' degrees of freedom  between source and drain reservoirs, cf. Fig. (1a). Detailed balance for the transition rates then already restricts the form of possible feedback models that can be obtained from microscopic interactions. Choosing a classical limit of weakly interacting uncharged fermions, we interpret the interaction--induced strong reduction of  fluctuations as a `condensation' in the space of  reservoir particle numbers. This is 
accompagnied with an instantaneously increased flow of information between a given channel and the rest of the system.
For periodic channel structures, Fourier analysis yields detailed predictions for the diffusion--like spreading of the autonomous feedback signal across the channels, which can be made visible in 
observables like the particle noise spectrum.

\section{Model}
Our model is defined by a $d$--dimensional lattice with $N$ lattice sites $\vecR_l$,  $l=1,...,N$, each of which serves as a connector within a channel with   
particle flow  from a source  to a drain reservoir at  rate $\gamma_{l+}$ (`forward'), and backwards from drain to source  at  rate $\gamma_{l-}$, cf. Fig. (1a).
There is no direct transfer of particles between different channels, but the time-independent rates $\gamma_{l\pm}=\gamma_{l\pm}(\vecn)$ are allowed to depend on the state $\vecn\equiv(n_1,...,n_N)^T$ of all the drain reservoirs. These are  defined by the numbers $n_l\in\mathbb{Z}$ of additional particles in drain reservoir $l$, when counting starts at time $t=0$. 

The dynamics is described by a Markovian rate equation for the probability $p(\vecn,t)$ of the drain reservoirs to be in state $\vecn$ at time $t$, 
\be
\dot{p}(\vecn,t) = \sum_{\vecn'}\left[ \Gamma(\vecn,\vecn') p(\vecn',t) -\Gamma(\vecn',\vecn) p(\vecn,t)\right],
\ee
where $\Gamma(\vecn',\vecn)\equiv \sum_{l=1}^N\sum_\pm\gamma_{l\pm}(\vecn) \delta_{\vecn',\vecn\pm \vece_l}$ 
with the Cartesian unit vector $\vece_l =(0,...,1,...0)^T$, such that we can  write
\be\label{pn}
\dot{p}(\vecn,t) =  \sum_{l=1,\pm}^N \left[ \gamma_{l\mp}(\vecn\pm\vece_l) p(\vecn\pm\vece_l,t) - \gamma_{l\mp}(\vecn) p(\vecn,t) \right],
\ee
which makes the contributions from forward and backward jumps in the $N$ channels explicit. 
As initial condition for $ p(\vecn,t)$  we set $p(\vecn,0)=\delta_{\vecn,0}$.

The rate equation \eq{pn} can be generalized to the more complex situation where the $N$ connectors between the source and drain reservoirs have themselves an internal structure, i.e. internal states (like in quantum dots) that can be occupied by particles. It seems difficult to make progress then (apart from large-scale numerics), but on the other hand
the above rate equation then can follow as an effective model  via  adiabatic elimination when time-scale separation is possible. For example, this is the case 
with one internal state connected to a source at a very fast rate and a drain at a very slow rate (not shown here).
In all what follows,  we will therefore work with  \eq{pn} as the starting point.

The rates $\gamma_{l\pm}(\vecn)$ are the key elements in describing the interactions among the channels and the effective feedback mechanism created thereby, which is why we discuss 
them in quite some detail in the following.

\begin{figure}[t]
\includegraphics[width=\columnwidth]{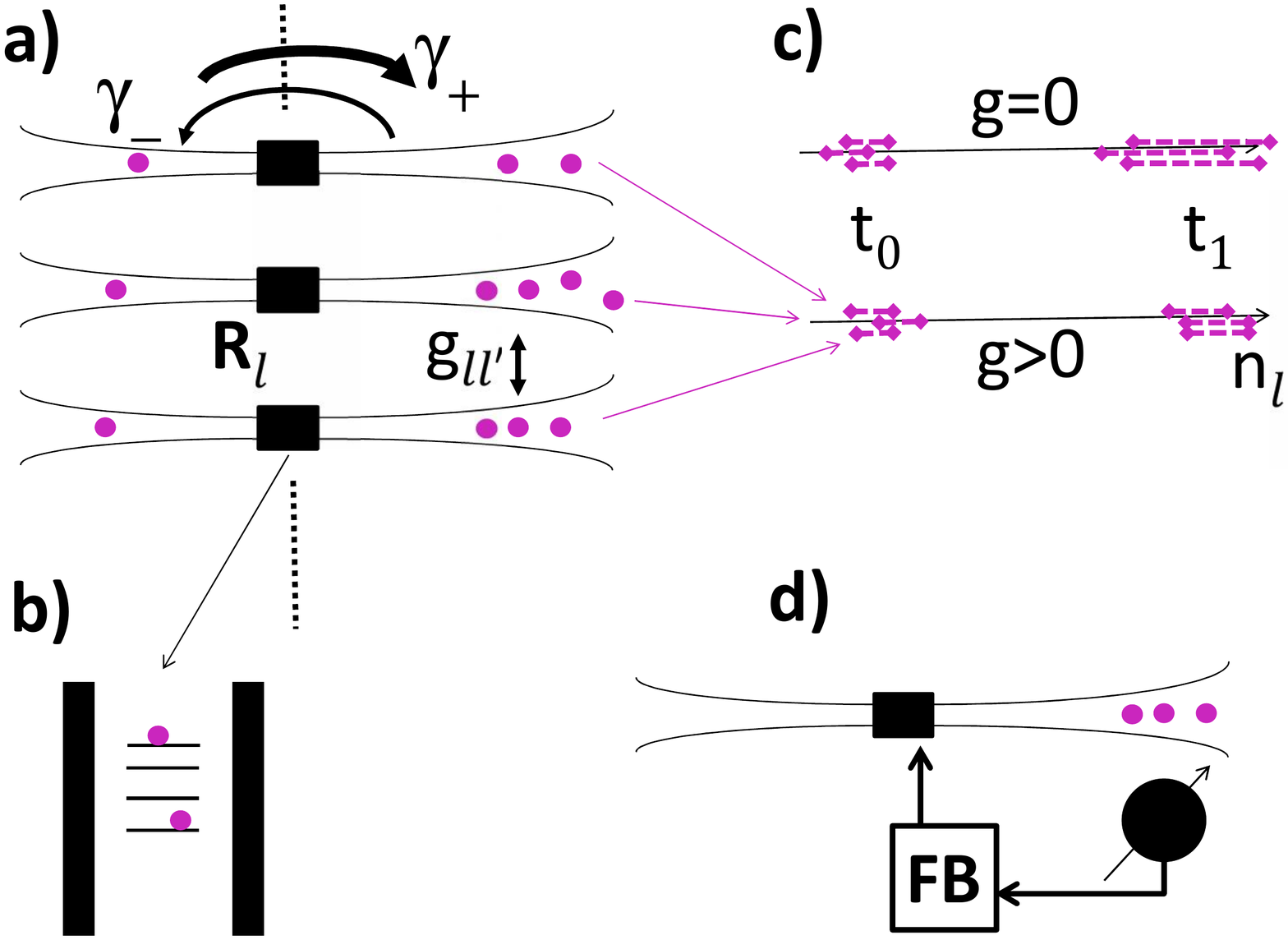}
\caption[]{\label{fig1}
\textbf{(a)} Particle transport between source and drain reservoirs at rates $\gamma_\pm$ through connectors in channels at positions $\vecR_l$. \textbf{(b)} Connectors could have internal states. \textbf{(c)} Autonomous feedback via interactions ($g>0$) among drain reservoirs reduces fluctuations of particle numbers $n_l$, represented as Brownian particles on a line. This is compared to the non-feedback case $g=0$ during time evolution from $t_0$ to $t_1$. \textbf{(d)} Active, non-autonomous measurement-based feedback for a single channel \cite{Bra10}.

}
\end{figure} 

\subsection{Decoupled channels with time-dependent rates} 
In the simplest version of the feedback model, the rates have the time ($t$) and $n$--dependent form
\be\label{gammasynch}
\gamma_{l\pm}(\vecn) =  f_\pm\left( I_l t-n_l\right),
\ee
where the function $f_\pm(x)$ describes a synchronization between the actual number $n_l$ with an external  reference $I_l t$ that grows linearly with time (`clock' with a fixed control current $I_l$). 
The form \eq{gammasynch} then immediately reduces  \eq{pn} to decoupled rate equations for  the single-channel  probabilities $p_1(n,t)$, cf. \eq{dotp1nta} in Appendix A. 
Physically, this kind of feedback can be  achieved  via an instantaneous control, i.e. a modification of the rates `by hand' in an active way  by an external agent like an electronic circuit, cf. Fig. 1d). 

For example, a linear modulation $f_\pm(x) = \gamma_\pm (1 \pm g x)$ with control parameter $0<g\ll 1$ continuously compensates particle number fluctuations in that the forward rate $\gamma_+$ is decreased  if there are too many particles as compared to a reference charge $I t$.
The rate equation  can then be solved exactly, and the main result \cite{Bra10} is a freezing of the single channel full counting statistics $p_1(n,t)$ at large times $t$ around a mean $\langle n \rangle_t \propto t$. Quantitatively, the second cumulant   follows from  \eq{pn}   as
\be\label{secondcumnonauto}
\langle n^2 \rangle_t - \langle n \rangle_t^2 = \frac{1}{2g} \left( 1- e^{-2 g (\gamma_++\gamma_-) t} \right),
\ee
which means that fluctuations are strongly suppressed for $g>0$,  which is schematically shown in  Fig. (1c).
In a similar way, expressions for higher cumulants \cite{Bra10} can be derived that all saturate at finite values $\propto 1/g$ at large times $t$.

\subsection{Interacting channels}
Alternatively,  feedback occurs via the design of interactions among particles. In most of what follows we will be dealing with an autonomous form 
\be\label{rates}
\gamma_{l\pm}(\vecn) = \gamma_\pm \left[1 \pm \sum_{j=1}^N g_{lj}(n_j-n_l)\right],
\ee
which is parametrized by interaction matrix elements $g_{lj}$ and which is linear in the $n_l$. The rates \eq{rates} are a generalization of the \eq{gammasynch} in the following way: consider the simplest all-to-all coupling $g_{ij}=\frac{g}{N}$ for which
the rates have the form
$
\gamma_{l\pm}(\vecn) =  f_\pm\left( n_{\rm av}-n_l\right),
$
with the average $  n_{\rm av}\equiv   (n_1+...+n_N)/N$ and 
$f_\pm(x) = \gamma_\pm (1 \pm g x)$ as above.
Now, for $N\to \infty$ the average $ n_{\rm av}$  becomes a macroscopic variable, and neglecting the 
fluctuations of the latter against those of $n_l$ one obtains the rates \eq{gammasynch} with $I_lt = \langle  n_{\rm av} \rangle $, i.e. 
we recover the synchronization model in this mean-field limit (see Appendix A for some more details). 

The rates \eq{rates} have a very intuitive form: 
the $n_l$--dependence leads to a dynamical compensation of particle number differences among the channels. If there are too many particles in the drain reservoir $l$ (as compared to all the other drains $j\ne l$), the forward (backwards) rate for channel $l$ is decreased (increased), and the other way round if there are too many particles in drain $l$. As we will show below, this compensation mechanism tends to suppress  {\em fluctuations} of the drain reservoir particle numbers $n_l$ also within each single channel $l$. 

The rates \eq{rates} can potentially become negative (and the model unphysical) in the linear approximation; however, this is not really a problem as long as the interactions are sufficiently small. More 
importantly,  with the  $g_{lj}$ as unspecified parameters, we have not yet addressed the issue of detailed balance between forward and backward rates so far.

\subsection{Detailed balance}
We therefore consider a dependence of  $\gamma_{l+}(\vecn)\equiv r_l(\varepsilon_l (\vecn)) $  on the energy differences
$\varepsilon_l (\vecn)\equiv\Delta\mu_l + \Delta E_l(\vecn)$,  where 
$\Delta\mu_l \equiv \mu_{l,{\rm d}} -\mu_{l,{\rm s}}$ is the difference of the chemical drain and source potentials in channel $l$,
and $ \Delta E_l(\vecn)\equiv E(\vecn + \vece_l)  - E(\vecn)$
 with the energy $E(\vecn)$ due to interactions among the particles in all drain channels. Correspondingly, $\gamma_{l-}(\vecn+\vece_l) = r_l(-\varepsilon_l (\vecn))$, 
and detailed balance reads
\be\label{db}
\gamma_{l+}(\vecn) = \gamma_{l-}(\vecn+\vece_l) e^{-\beta\left[\Delta\mu_l + \Delta E_l(\vecn)\right] },
\ee
where all the channels are kept at the same temperature $k_BT=\beta^{-1}$.

By expanding the rates  $r_l(\varepsilon_l (\vecn))$ and the r.h.s. of \eq{db} in the interaction energies $\Delta E_l(\vecn)$, we find a linearization
\be\label{rates2}
\gamma_{l\pm}(\vecn) = \gamma_{l\pm}\times \left( 1 + \frac{r_l'(\pm \Delta\mu_l )}{r_l(\pm \Delta\mu_l )} \left[ E(\vecn \pm \vece_l) -  E(\vecn)\right]   \right)
\ee
with $\gamma_{l\pm} \equiv   r_l(\pm  \Delta\mu_l)$, which is consistent with detailed balance if $\Delta E_l(\vecn)$ is sufficiently small.

We can be more specific in \eq{rates} if we model the particles as fermions that tunnel between a non-interacting  (source) and an interacting (drain) region, similar to the 
orthodox model of Coulomb blockade for charged fermions (electrons) in metallic quantum dots \cite{BruusFlensberg}. In this case, the specific expression 
\be\label{fermionrates}
r_l(\varepsilon) = r_l \frac{\varepsilon}{e^{\beta \varepsilon}-1}
\ee
can be derived, where $r_l$ is a parameter  that depends on the (flat) density of states in source and drain reservoirs.  The coefficient in the rate \eq{rates2} then becomes  $\frac{r_l'(\pm \Delta\mu_l )}{r_l(\pm \Delta\mu_l )}\approx -\beta/2$ provided
$\beta |\Delta \mu_l| \ll 1$,
in which case detailed balance is always fulfilled in lowest order in $\beta \Delta E_l(\vecn)$ in \eq{db}. This also shows that in this expansion, we are dealing with rates in a high-temperature limit, where the particles essentially behave as classical objects and quantum statistics and quantum coherence plays no role. What we are left with as a key modeling parameter are the interactions among the particles.

\subsection{Interaction form}
In order  to make analytical progress with \eq{pn}, we further linearize the rates $\gamma_{l\pm}(\vecn)$ in $\vecn$
by assuming a quadratic dependence 
\be\label{Equad}
E(\vecn) \equiv \frac{1}{2}\vecn^T U \vecn
\ee
with a symmetric $N\times N$ matrix $U$ of interaction parameters whence \eq{rates2} becomes
\be\label{rates1}
\gamma_{l\pm}(\vecn) &=& \gamma_{l\pm}\left[1  \mp \frac{\beta}{2}\vece_l^T U \vecn\right],
\ee
where we neglected a constant, $\vecn$--independent  term.

From \eq{rates1} we recognize that we obtain the feedback form \eq{rates} with the identification
\be\label{Uij}
U_{lj} = 
\frac{2}{\beta}  \left( \delta_{lj}\sum_{i=1}^N g_{li}  -  g_{lj}\right)
\ee
and homogeneous rates $\gamma_{l\pm} = \gamma_\pm$ independent of channel index $l$. 
For example, with $g_{lj}=\frac{g}{N}$ and a single interaction parameter $g>0$, one 
has $U_{lj} = \frac{2}{\beta}g  (\delta_{lj} -\frac{1}{N})$, which mimics a mean--field type repulsive interaction within the same channel $l=j$, and an attractive interaction between  particles in different channels $l\ne j$, regardless of particle distance. 
The form \eq{Uij} means that the interaction matrix $U$ has a zero eigenvector $(1,1,...,1)^T$ such that 
\be\label{zerocond}
\sum_{j=1}^NU_{lj}=0
\ee
and the (repulsive)  interaction strength $U_{ll}$ within each channel $l$ is cancelled by the sum of the remaining (attractive) interactions with all other channels $j$.
Physically, such an interaction corresponds to a repulsive short-range and an attractive  long-range, van-der-Waals like force between uncharged particles.

The choice of parameters in the interaction matrix $U_{lj}$ in \eq{Equad} is restricted to the particular form \eq{Uij}, because we demand a stationary particle current to flow through the channels, i.e. a situation where the average numbers $n_l$ continue to grow linearly with time $t$. 

We obtain the equation of motion for these averages $\langle n_l \rangle\equiv \sum_\vecn n_l p(\vecn,t)$ from \eq{pn} by summation over $n_l$, as 
\be\label{1stmoment}
\frac{d}{dt}\langle n_l \rangle = \langle \gamma_{l+}(\vecn) -  \gamma_{l-}(\vecn) \rangle
\ee
or
$
\frac{d}{dt}\langle \vecn \rangle = \Delta \vecgamma + (G^++G^-) \langle \vecn \rangle,
$
in matrix form 
with the matrix $G_{lj}^\pm \equiv -\frac{\beta}{2}\gamma_{l\pm} U_{lj}$ and the vector $\Delta \vecgamma_l = \gamma_{l+}-\gamma_{l-}$. For general $U_{lj}$, this equation has a fixed point with $\langle \vecn \rangle=-G^{-1}\Delta \vecgamma  $ for invertible $G\equiv G^++G^{-1}$. Here, however, we are interested in the opposite case where $G$ (and thus $U$) is not regular but has a zero eigenvector $(1,1,...,1)^T$ leading to \eq{zerocond} and \eq{Uij}. 

As an immediate consequence, we find the solution of \eq{1stmoment}
\be\label{1stmomentresult}
\langle n_l \rangle = (\gamma_+ - \gamma_-) t,
\ee
i.e. a linear increase of the particle numbers in all channels regardless of the interaction strength.

\section{Fluctuations}
The average flow of particles has the rather trivial form \eq{1stmomentresult} for the homogeneous case considered here and in the following, i.e. equivalence of all channels $l$ with translational invariance across the $d$--dimensional lattice of connectors between source and drain reservoirs. As in our previous work \cite{Bra10}, we expect the feedback to drastically modify the full counting  statistics of the particle numbers $n_l$. The most important object to quantity fluctuations then is the second cumulant. Higher cumulants are difficult to extract from \eq{pn}, except for special cases as in \cite{Bra10}, and we will therefore effectively evaluate only a Gaussian model for the fluctuations.

\subsection{Second cumulant}
We quantify the fluctuations by the correlation function
\be\label{cumulantdef}
C_{ll'}(t)\equiv \langle n_l n_{l'}\rangle - \langle n_l\rangle\langle n_{l'}\rangle,
\ee
for which we obtain an equation of motion via \eq{pn}. Multiplication of \eq{pn} with $(\vecn \vece_l)(\vecn \vece_{l'})$ and summation over $\vecn$ yields $\frac{d}{dt}\langle n_l n_{l'}\rangle = \sum_\pm \langle \delta_{ll'} \gamma_{l\pm}(\vecn) \mp n_l \gamma_{l'\pm}(\vecn) +(l\leftrightarrow l') $, and for the homogeneous case with rates \eq{rates} we obtain 
\be\label{Cdiffequation}
\frac{d}{dt}C_{ll'} = \gamma\left(\delta_{ll'} +\sum_{j=1}^Ng_{l'j}\left( C_{lj}- C_{ll'}   \right) + l\leftrightarrow l'  \right)
\ee
with the definition $\gamma\equiv \gamma_+ + \gamma_-$.
We can now take advantage of the discrete translational invariance on the lattice $\{\vecR_l \}$ of channel connectors $l$ by defining the Fourier decomposition $C_{ll'} =\frac{1}{N}\sum_{\veck\veck'} \hat{C}_{\veck\veck'}e^{-i\veck\vecR_l + i\veck'\vecR_{l'}}$ in reciprocal space for the correlation function and 
\be\label{gljdef}
g_{lj} = \frac{1}{N}\sum_{\veck} e^{-i\veck(\vecR_l -\vecR_j)} \hat{g}(\veck)
\ee
for the interaction matrix elements in \eq{rates}. 

We solve \eq{Cdiffequation} for an initially empty  drain reservoirs with $C_{ll'}(0)=0 $, 
\be\label{Clj}
C_{ll'}(t) = \frac{1}{N}\sum_{\veck} e^{-i\veck(\vecR_l -\vecR_{l'})} \sigma_\veck(t),
\ee
with the definitions 
\be\label{Gammarates}
 \sigma_\veck(t)\equiv \frac{\gamma}{\Gamma_\veck}\left( 1- e^{-\Gamma_\veck t}\right),\quad
\Gamma_\veck \equiv 2 \gamma (\hat{g}(0)-\hat{g}(\veck)),
\ee
which are real quantities as we recognize from the representation $ \hat{g}(\veck) = \frac{1}{N} \sum_{ll'}  e^{i\veck(\vecR_l -\vecR_{l'})} g_{ll'}$ and the symmetry $g_{ll'}=g_{l'l}$.

{\em Infinite range interaction .---} 
We first evaluate the result \eq{Clj} for the mean--field type  model $g_{ij}=\frac{g}{N}$ in the interactions \eq{Uij}, which (simple as it is) already contains the essential ingredients of the feedback mechanism we try to analyse here. 

Using $\hat{g}(\veck) =  \frac{g}{N^2} \sum_{ll'}  e^{i\veck(\vecR_l -\vecR_{l'})} = g \delta_{\veck,0}$, from \eq{Gammarates} we obtain
\be\label{infiniterange}
\Gamma_{\veck\ne 0}=2g\gamma.
\ee
The $\veck=0$ contribution in the sum \eq{Clj} is extracted  with  the limit 
$\Gamma_{\veck=0}=0$, and we  have
\be\label{CLLMF}
C_{ll'}(t) &=& \frac{\gamma t}{N} + \left(\delta_{ll'} -\frac{1}{N}\right)     \frac{1}{2g}\left(1-e^{-2\gamma g t} \right),
\ee
where we used $\sum_{\veck\ne 0}  e^{-i\veck(\vecR_l -\vecR_{l'})} = N\delta_{ll'} -1$ for the second term. 

The result \eq{CLLMF} already captures one of the main ingredients of the feedback mechanism described by our model: at large times $t$, the fluctuations $C_{ll'}(t)$ increase linearly but are suppressed by a factor of $1/N$ as compared to the simple non--interacting case with the Poissonian fluctuations $\langle \delta n^2_l(t)\rangle =\gamma t$. 

\begin{figure}[t]
\includegraphics[width=\columnwidth]{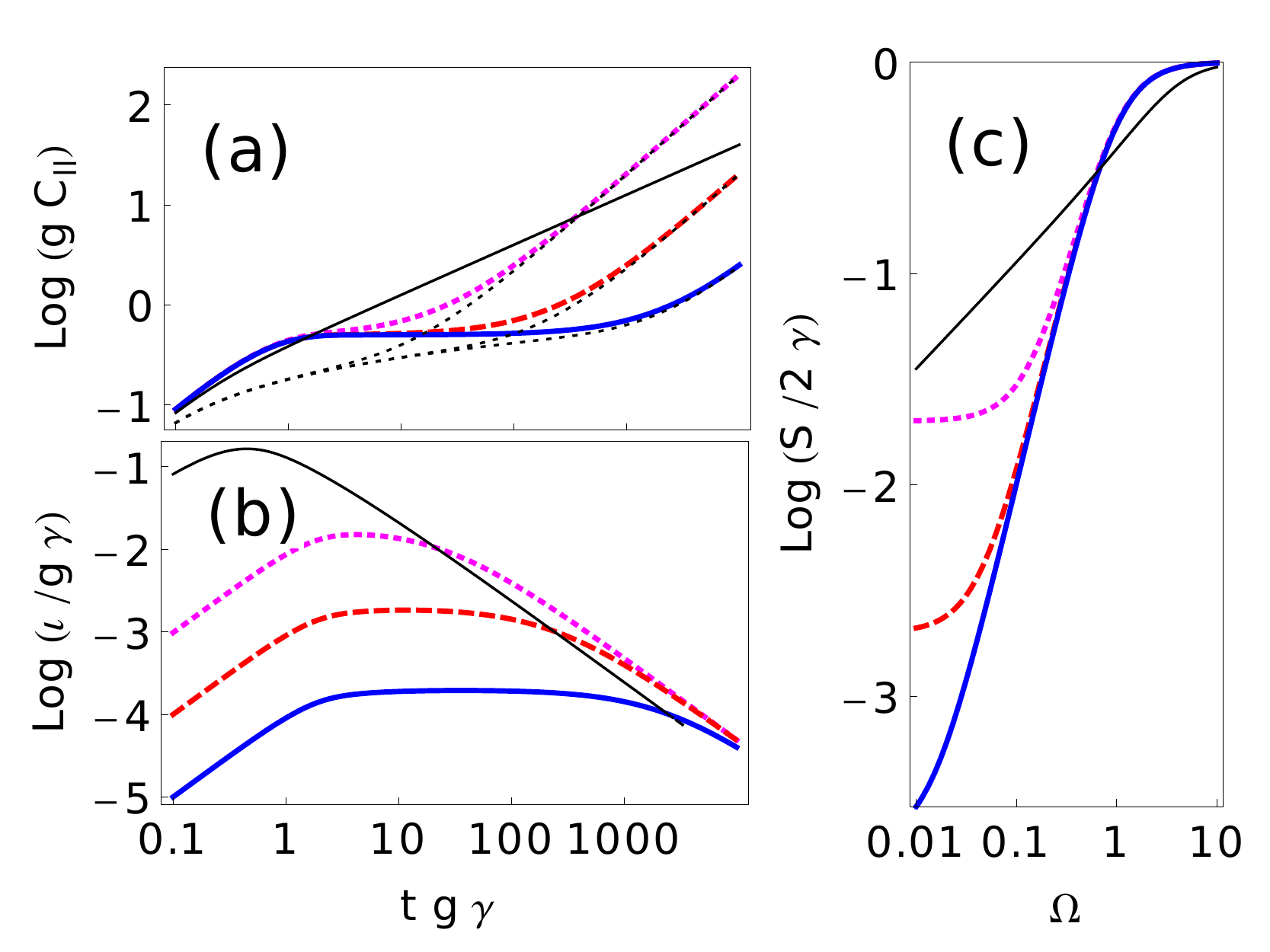}
\caption[]{\label{cumulantinfo}
\textbf{(a)} Diagonal second cumulant $C_{ll}$ \eq{CLLMF} multiplied with  interaction parameter $g$ and \textbf{(b)} information current $\iota_{1'\to 1}$, \eq{info}, as a function of time $t$ multiplied with $g\gamma$. 
\textbf{(c)} Stationary diagonal noise spectrum ($l=l'$, \eq{Snoise}) as a function of scaled frequency $\Omega\equiv  \omega/(g\gamma)$.
Curves for infinite range interaction model \eq{infiniterange} with channel numbers $N=50$ (magenta dotted), $N=500$ (red dashed), $N=5000$ (blue thick line), 
diffusive approximation \eq{difflimit} in $d=2$ dimensions with finite $N$ in (a), and nearest neighbor model \eq{nnmodel} in  $d=1$ dimension with $N\to \infty$ (thin black line).}
\end{figure}

In the limit of a very large number of channels, for times such that $N \gg \gamma t \gg 1$ we obtain a freezing of the fluctuations towards the value  $C_{ll'}(t) \to \delta_{ll'}/{2g}$. For the  fluctuations within one channel, this becomes visible as a  plateau at intermediate times in Fig. (\ref{cumulantinfo}a), where we plot $C_{ll}(t)$ for various values of $N$. 

In each channel $l$, a synchronization occurs where in the rates \eq{rates} the contributions from all the other channels can be replaced by a time-dependent average \eq{gammasynch} with $I_lt =(\gamma_+-\gamma_-)t$.
This is in accordance \cite{Bra10} with the non-autonomous model \eq{gammasynch}, where we  found a freezing of the entire full counting statistics around a moving mean value \eq{1stmomentresult} at large times $t$, with a second cumulant \eq{secondcumnonauto} that co--incides with  \eq{CLLMF} for $N\to \infty$. In our mean--field model here, instead of synchronization with an external `clock' current $I_l$, we have auto--synchronization between all channels. 
In Appendix A, we demonstrate the equivalence between the synchronization model \eq{gammasynch} and the $N$--channel model with infinite range interactions in the mean field limit  by deriving the equation of motion for the single channel probabibilty $p_1(n,t)$. 

{\em Diffusion limit and scaling.---}
We now evaluate the result \eq{Clj} in the opposite limit of short--range interactions $g_{ij}$, which has a more interesting dynamics as compared to the mean--field case above. We can evaluate the Fourier--matrix elements $\hat{g}(\veck)$ for nearest--neighbor interaction in \eq{gljdef}, 
\be\label{nnmodel}
g_{lj}= g \delta_{\langle lj\rangle},\quad g>0.
\ee
For example,  on an infinite  lattice with lattice constant $a$ in $d=1$ dimension, we have $\hat{g}(k) =  2g \cos ka$.

The decay rates \eq{Gammarates} acquire the form
\be\label{difflimit}
\Gamma_\veck = 2\gamma g a^2 \veck^2,\quad |\veck|\to 0
\ee
in the continuum limit where the connector lattice constant $a$ approaches zero. In the terminology of dynamical phase transitions, 
this corresponds to  model--A dynamics \cite{HH77,KardarBook} due to the fact that the total particle number $N_{\rm tot}\equiv \sum_ln_l$  in the drains grows with time and is not conserved.  

The diffusive nature of the feedback dynamics is now  clear from the second cumulant \eq{Clj}, the 
temporal derivative of which, 
\be
\dot{C}_{ll'}(t) = \frac{\gamma}{N}\sum_{\veck} e^{-i\veck(\vecR_l -\vecR_{l'})}  e^{-2D \veck^2 t},
\ee
has the form of a $d$--dimensional diffusion propagator, where $D\equiv \gamma g a^2 $ is an effective diffusion constant. Accordingly, we obtain the simple,  $d$--dependent temporal and spatial (with respect to $|\vecR_l -\vecR_{l'}|$) scaling behavior of ${C}_{ll'}(t)$ characteristic of a critical dissipative Gaussian model \cite{KardarBook} with dynamical critical exponent $z=2$.

For any finite number $N$ of channels, the integral approximation to the $\veck$--sum \eq{Clj} has a lower cutoff at the inverse system size $\propto N^{-1/d}$. With increasing $d$, 
at intermediate times $C_{ll}(t)$ develops a flat plateau as in the all-to-all coupling model \eq{infiniterange}, before  the linear in $t$ behaviour sets in at very large times. 
This is shown in Fig. (\ref{cumulantinfo}a) for the diffusive model \eq{difflimit} in $d=2$ (with the same values for $N$ as in the all-to-all coupling model), 
and for the full nearest--neighbor interaction model \eq{nnmodel} in $d=1$ in the limit $N=\infty$, where $C_{ll}(t\to\infty) \propto t^{1/2}$.

\subsection{Noise Spectrum}
The above temporal fluctuations can be related to the fluctuations $\delta I_l(\tau)\equiv I_l(\tau)-\langle I_l \rangle$ of the particle currents $I_l(\tau)$ in a stationary situation, as quantified by the symmetrized noise spectrum 
\be\label{Snoise}
S_{ll'}(\omega)\equiv 
\int_{-\infty}^\infty d\tau e^{i\omega \tau} \langle  \delta I_l(\tau)  \delta I_{l'}(0) +  \delta I_{l'}(\tau)  \delta I_{l}(0)   \rangle. 
\ee
We use  the MacDonald formula \cite{MacDonald,LAB07} together with the regression theorem  \cite{Carmichael_02} to obtain the two-time correlation functions
$\left\langle \delta n_l(t) \delta n_{l'}(t+\tau)\right\rangle$
for $t\to\infty$ (cf. Appendix C) and find
\be\label{noisefinal}
S_{ll'}(\omega) = 2\frac{\gamma}{N}\left( 1+ \omega^2\sum_{\veck \ne 0} \frac{\cos \left({\veck(\vecR_l-\vecR_{l'}) }\right)}{\Gamma_\veck^2/4 +\omega^2}\right).
\ee
We have evaluated \eq{noisefinal} for the infinite range interaction model, \eq{infiniterange}, and for the nearest-neighbor interaction model \eq{nnmodel}  in $d=1$ dimension. In both cases, the diagonal noise $l=l'$ is independent of the channel index $l$ and displays scaling , i.e. a dependence on the dimensionless variable $\Omega\equiv \omega/(g\gamma)$ that in both cases contains the feedback strength parameter $g$,
\be
\frac{S_\infty}{2\gamma} &=& \frac{1}{N} + \frac{N-1}{N} \frac{\Omega^2}{1+\Omega^2 }\label{noiseexamples1}\\
\frac{S_{\rm d=1}}{2\gamma} &=& \sqrt{ \frac{1}{2} \left( \frac{1}{\sqrt{ 1+16/\Omega^2} } +  \frac{1}{ 1+16/\Omega^2 }\right)}\label{noiseexamples2}.
\ee
The noise spectra in these two cases have the correct high-frequency (or zero feedback $g=0$) Poissonian limit $S(\Omega\to \infty)=2\gamma$, cf. Fig. (\ref{cumulantinfo}c).

The opposite limit of zero frequency $\omega=0$ corresponds to the Poissonian long--time $t\to \infty$ limit \eq{CLLMF} at any finite $N$ for $S_\infty$, whereas the $N=\infty$ lattice model (nearest neighbor interaction) is determined by diffusive dynamics in $d$ dimensions at small frequencies (that fails on short time- or large frequency scales),
\be
\frac{S_{\rm d}(\omega)}{2\gamma}= \frac{1}{2\sqrt{2}} \left( \frac{\omega}{g\gamma} \right)^{\frac{d}{2}},\quad \omega \to 0,
\ee
which also follows from \eq{difflimit} with \eq{noisefinal}.
The noise reduction at low frequencies highlights the strong suppression of fluctuations in each transport channel due to the feedback from all the other channels.

\section{Heat and information}

\subsection{Condensation picture}
The interaction--induced feedback mechanism among the channels leads to a `condensation' of the system into a state $p(\vecn,t)$ where fluctuations of the particle numbers $n_l$ are essentially suppressed at large times.

Without feedback, the addition of the $N$ independent Poissonian processes (channels) in the {\em total} particle number $n_{\rm tot}\equiv \frac{1}{N}\sum_ln_l$ (normalized by $N$) 
leads to $\langle \delta n_{\rm tot}(t)^2\rangle =  \gamma t/N  $: the fluctuations of the macroscopic variable $n_{\rm tot}$ are reduced by a factor $1/N$ as compared to the 
fluctuations $\langle \delta n_l(t)^2\rangle =  \gamma t $ in the individual, `microscopic' channel variables $n_l$. 

In contrast, with feedback the microscopic fluctuations are reduced at large times $t$, $\langle \delta n_l(t)^2\rangle \to  \gamma t/N $, whereas the form \eq{Gammarates} guarantees that the fluctuations of  $n_{\rm tot}$ are not affected by the feedback: summation of \eq{Clj} over $l$, use of $\frac{1}{N} \sum_{ll'}  e^{-i\veck(\vecR_l -\vecR_{l'})} = N \delta_{\veck,0}$ and the limit $\Gamma_{\veck=0}=0$ again leads to $\langle \delta n_{\rm tot}(t)^2\rangle =  \gamma t/N$, and the microscopic and macroscopic fluctuations are now of the same order. 

This explains the scheme shown in Fig. (1c), where the channel variables $n_l$ are (continuous)  position coordinates  $x_l$ of a large cluster of   $N$ interacting Brownian particles on a line, with the cluster center--of--mass coordinate corresponding to $ n_{\rm tot}(t)$. The feedback (interactions) then condenses the (at time $t=0$) `gas'  of initially independent particles into a tightly bound cluster, where fluctuations of {\em relative} distances between particles do not grow with time (as without feedback) but approach the fixed value obtained from \eq{Clj} as
\be
\langle (n_l -n_{l'})^2 \rangle_{t\to \infty} =  \frac{2 \gamma }{N}\sum_{\veck\ne 0} \frac{ 1 - \cos \left(\veck(\vecR_l -\vecR_{l'})\right)}{\Gamma_\veck}.
\ee

\subsection{Heat}
Starting from initially empty drains, this condensation process is accompagnied by an increase in interaction energy $E(\vecn)$, \eq{Equad}, which has to be compensated by a flow of heat from the reservoirs that keep the drains at constant temperature and chemical potential all the time. Explicitly, a simple calculation leads to 
\be\label{heat}
\frac{d}{dt}\langle E(\vecn) \rangle_t = \frac{1}{2\beta} \sum_\veck \Gamma_\veck e^{-\Gamma_\veck t}, 
\ee 
with the rates $\Gamma_\veck$ defined in \eq{Gammarates}. By integration of \eq{heat}, the quadratic interaction potential \eq{Equad} thus leads to the equipartition
$\langle E(\vecn) \rangle_\infty - \langle E(\vecn) \rangle_0 = \frac{1}{2} k_BT N$, i. e. an interaction energy $\frac{1}{2} k_BT$ stored in each channel.

At the same time, the constant  current $I\equiv \gamma_+ -\gamma_-\ge 0$ of particles (cf. \eq{1stmomentresult}) generates a dissipative power $-I \Delta\mu\equiv I (\mu_{{\rm s}} -\mu_{{\rm d}}) \ge 0$ 
in each channel that has to be dissipated into the reservoirs, and the total amount of heat flow $\dot{Q}$ into the reservoirs thus is  
\be\label{Qdotsimple}
\dot{Q}\equiv -N I \Delta\mu - \frac{d}{dt}\langle E(\vecn) \rangle_t.
\ee
We note that this simple splitting of $\dot{Q}$ into two contributions is a consequence of our model for the rates \eq{rates1},  with its separation of excitation energies 
$\varepsilon_l (\vecn)\equiv\Delta\mu_l + \Delta E_l(\vecn)$ into a constant single particle part $\Delta\mu_l$ and the part $ \Delta E_l(\vecn)$ with the interaction energies \eq{Equad}.

In the following, we check \eq{Qdotsimple} by a somewhat more elaborate argument based on entropy.

\subsection{Fokker--Planck equation}
We use an approximation to the original rate equations \eq{pn} where the integers $n_l$ become continuous, real number $x_l$ of which only Gaussian fluctuations are kept. In this way, we obtain a  Fokker--Planck equation for the probability distribution $P(\vecx,t)$ in the usual form of a continuity equation 
\be\label{FP}
\frac{\partial}{\partial t} P(\vecx,t) &=& -\sum_{l=1}^N \frac{\partial}{\partial x_l} J_l(\vecx,t)\\
 J_l(\vecx,t) &\equiv&  F_l(\vecx) P(\vecx,t) - \frac{\gamma}{2} \frac{\partial}{\partial x_l} P(\vecx,t),
\ee
with diffusion constant $\frac{\gamma}{2}$ and force term $ F_l(\vecx)  \equiv \gamma_+ - \gamma_- + \sum_{j=1}^N g_{lj}(x_j-x_l)$.

Since  $ F_l(\vecx)$ is linear, this can be solved exactly and thus be used to derive explicit expressions for the Shannon entropy 
\be\label{Shannon}
S(t) \equiv - \langle \ln P \rangle_t \equiv -\int d\vecx P(\vecx,t) \ln P(\vecx,t). 
\ee
Moreover, \eq{FP} is very convenient for directly reading off the splitting of the temporal change,
\be\label{Ssplitting}
\frac{d}{dt} S(t) = \Pi(t) - \Phi(t),
\ee
into a (positive) entropy production rate $\Pi(t)$ and the global entropy flow
\be\label{Phidef}
\Phi(t) \equiv \sum_{l=1}^N \left\langle \frac{2}{\gamma}  F_l^2(\vecx) + \frac{\partial F_l(\vecx) }{\partial x_l} \right\rangle_t
\ee
from the total system into the reservoirs \cite{Tom2006}. Crucially, even though $S(t)$ and $\Pi(t)$ are `abstract' quantities in the sense that they are no direct experimental observables, the entropy flow  $\Phi(t)$
directly yields  the (measurable) heat flow
$\dot{Q}\equiv \frac{1}{\beta}\Phi(t) $  into the reservoirs. The expectation value \eq{Phidef} can be expressed in terms of the second cumulants $\langle \delta x_l \delta x_{l'}\rangle $
via the identification $n_l \leftrightarrow x_l$ from  the $C_{ll'}(t)$, \eq{cumulantdef}, or from the direct solution of \eq{FP} (see below). Using  Fourier transformation and some straightforward algebra,
one obtains
\be\label{QdotFP}
\dot{Q}\equiv \frac{1}{\beta}\Phi(t)  = -\frac{2N I}{\beta} \tanh \frac{\beta\Delta\mu}{2} -  \frac{d}{dt}\langle E(\vecn) \rangle_t.
\ee
Now recalling that in the derivation of our classical transitions rates after \eq{fermionrates} we demanded $\beta |\Delta \mu| \ll 1$, we may replace the $\tanh$ in \eq{QdotFP} by its argument and thus recover our previous result \eq{Qdotsimple} that was based upon a simple energy argument.

\subsection{Feedback information flow}
Finally, we quantify the information flow in the `feedback--freezing' of the distribution $P(\vecx,t)$. \eq{FP} models the continuous counting variables $x_l$ as the positions of $N$ linearly coupled 
Brownian particles, a situation analysed in detail (for $N=2$) by Allahverdyan {\em et al.} \cite{AJM2009}  via the mutual information $\mathcal{I}$. Recently, Horowitz \cite{Hor2015} has generalized 
this to multipartite systems governed by Fokker--Planck equations like \eq{FP} and introduced an information flow based on the concept of neighbours influencing each other. 

Here, we use the simplest bipartite splitting of the whole system into a single channel $l=1$ (with probability density $P_1(x_1,t)$) and the remaining $N-1$ channels (with probability density $P_{1'}(x_2,...,x_n,t)$), where the mutual information 
\be\label{infodef}
\mathcal{I}(t)\equiv \int d^N\vecx P(\vecx,t) \ln \frac{P(\vecx,t)}{P_1(x_1,t)P_{1'}(x_2,...,x_n,t)    }
\ee
quantifies the build-up of feedback effects within a  small subsystem (single channel)  embedded into a large feedback environment. 

The temporal change $\dot{\mathcal{I}}(t)=\iota_{1'\to 1}+ \iota_{1\to 1'}$ defines the two information flows between the two subsystems \cite{AJM2009,HE2014}. 
Using \eq{FP} and the equivalence of all channels, we can express the flows via entropies \cite{AJM2009}, such as 
\be\label{info}
\iota_{1'\to 1}(t) = \dot{S}_1(t) - \frac{1}{N} \dot{S}(t),
\ee
with the Shannon entropy $S(t)$ of the total system, \eq{Shannon}, and the quantity
\be\label{S1def} 
\dot{S}_1(t)\equiv \int d^N\vecx \left( \frac{\partial}{\partial x_1} J_1(\vecx,t) \right)  \ln  P_1(x_1,t) 
\ee
which has the form of a  {\em local} Shannon entropy change, cf. Appendix C. 

In $\veck$--space, we find that the second cumulants $\sigma_\veck(t)$ (defined in \eq{Gammarates}) completely determine the entropies via
\be\label{S1Sexplicit}
 \dot{S}_1(t) = \frac{1}{2}\frac{d}{dt} \ln \sum_\veck \sigma_\veck(t),\quad \dot{S}(t) = \frac{1}{2}\frac{d}{dt}  \sum_\veck \ln \sigma_\veck(t),
\ee
which in view of our Gaussian approximation is not so surprising, but demonstrates that there is a direct connection between the (abstract) information flow \eq{info} and a fluctuation quantity that at least in principle is  accessible  via the counting statistics. This argument is in line with  recent results by Ansari and  Nazarov \cite{AN2015}, who derived a general connection between Renyi (and Shannon) entropy and full counting statistics.

At small times $t$, owing to $ \sigma_\veck(t\to 0) \approx \gamma t$, the information flow is zero although both entropy changes are large, $ \dot{S}_1(t) = \frac{1}{N} \dot{S}(t) \approx \frac{1}{2t}$, the latter being due to the initial condition $P(\vecx,t=0)=\delta^N(\vecx)$. On the other hand, at any finite $N$ and large times $t$, only the $\veck=0$ component significantly contributes to \eq{S1Sexplicit} since $ \sigma_\veck(t\to \infty)  \approx \delta_{\veck,0} \gamma t$ and the information flow decays towards zero again, $\iota_{1'\to 1}(t\to \infty) \sim \frac{1}{t} $.

In its transient dynamics, thus, $\iota_{1'\to 1}(t)$ has a maximum at some finite time of the order $g \gamma$, where $g$ is the dimensionless coupling constant of the interaction model. This behavior is shown in Fig. (\ref{cumulantinfo}).

For the infinite-range interaction model \eq{infiniterange}, the difference between local and total Shannon entropy vanishes with increasing $N$ since the notion of `boundary' between subsystems becomes meaningless. Accordingly, with $ \dot{S}_1(t) = \frac{1}{N} \dot{S}(t) = g(e^{-2g\gamma t}-1)^{-1}$ for $N\to \infty$, the information flow vanishes. On the other hand, at large but finite $N$, an interesting feature then is the plateau  in $\iota_{1'\to 1}(t)$  at intermediate times, which is shown  in the lower part of Fig. (\ref{cumulantinfo}) and which clearly corresponds to the  `freezing' of the fluctuations, i.e. the plateau in the second cumulants $C_{ll}(t)$.

\section{Discussion}
The  autonomous feedback control discussed here is based on a microscopic mechanism, but it comes with the constraints \eq{Uij} and \eq{zerocond}  for the interactions $U_{lj}$, which suggests uncharged particles as appropriate candidates, such as  ultracold atoms used in recent transport experiments \cite{ultracold}. In contrast, for charged particles (e.g., for electronic transport), 
the non-autonomous, synchronization feedback model \eq{gammasynch} (which essentially is an open-loop control scheme in the space of particle numbers $n$), appears to be more flexible. 

Needless to say that the choice of a particular control scheme depends on the specific control task, which in our model  was to stabilize the full counting statistics at a given particle current $I=\gamma_+-\gamma_-$ over a longer time $t$. One aspect then is the issue of energy cost and efficiency. The two control schemes mentioned above both scale linearly in $t$ in terms of energy costs: the  non-autonomous scheme by requiring a continuing  modulation of energy barriers, and the autonomous scheme by needing large channel numbers $N\propto t$ (and thus a large total heat flow $\frac{1}{2}k_BT N\propto t$, cf. after \eq{heat}) if stabilization over large intervals $t$ is required, as we saw in  Fig. (2a). 

The analysis proposed here remained at the level of a Gaussian approximation for the full counting statistics, i.e. a linear Fokker-Planck equation. Our results might stimulate further work to explore the connection between the recently discussed flow of information in feedback networks \cite{IS2013,HE2014,HBS2014,Hor2015}, and the  dynamical scaling type analysis for transport processes in interacting systems, such as in the Kardar-Parisi-Zhang (KPZ) model \cite{KardarBook} for particle deposition and surface growth.
For this latter point, the most urgent step then is to go beyond the linear (in $\vecn$) approximation \eq{rates1} of the rates, either by going to higher orders in the expansion \eq{rates2} (i.e., beyond the classical limit and towards lower temperatures), or by sticking to the classical limit  \eq{rates2} and employing interaction models beyond the simple quadratic model \eq{Equad}.
An interesting alternatively there could be the study of the KPZ equation (or other Langevin equations belonging to \eq{pn}) {\em per se} from the perspective of entropy and information flow.

\section*{Acknowledgements}
I thank M. Esposito, G. Schaller and P. Strasberg for valuable discussions, and I acknowledge support by the  DFG  via projects BR 1528/7-1, 1528/8-1, 1528/9-1, SFB 910 and GRK 1558.

\begin{appendix}
\section{Non-autonomous feedback in single-channel model}
Here, we derive the equation of motion for the reduced probability 
\be
p_1(n,t)\equiv \sum_\vecn  \delta_{n_1,n} p(\vecn,t), 
\ee
with $\vecn\equiv (n_1,...,n_N)^T$,
for a single channel (chosen as $l=1$) in the $N\to\infty$ limit of the constant interaction model in \eq{infiniterange}. 
After multiplying \eq{pn} with $\delta_{n_1,n}$ and summing over $\vecn$, only the $l=1$ term of the sum over $l$ remains and 
\be\label{dotp1nt}
&& \dot{p}_1(n,t) \\
&=&  \sum_{\vecn,\pm}\delta_{n_1,n} \left[ \gamma_{1\mp}(\vecn\pm\vece_1) p(\vecn\pm\vece_1,t) - \gamma_{1\mp}(\vecn) p(\vecn,t) \right].\nonumber
\ee
For $g_{ij}=\frac{g}{N}$, the rates have the form
$
\gamma_{1\pm}(\vecn) =  f_\pm\left( n_{\rm av}-n_1\right),
$
with the average $  n_{\rm av}\equiv   (n_1+...+n_N)/N$ and 
$f_\pm(x) = \gamma_\pm (1 \pm g x)$ for the linear rate model \eq{rates} with feedback parameter $g$.
Now, for $N\to \infty$ the average $ n_{\rm av}$  becomes a macroscopic variable. We neglect the fluctuations of the latter against those of $n_1$ by factorizing
\be
 \sum_\vecn  \delta_{n_1,n}\gamma_{1\mp}(\vecn) p(\vecn,t) &=& \sum_\vecn  \delta_{n_1,n} f_\mp\left( n_{\rm av}- n\right)  p(\vecn,t) \nonumber\\
&\approx & 
f_\mp\left( \langle n_{\rm av} \rangle_t- n\right)p_1(n,t)
\ee
and similar for the terms $ \gamma_{1\mp}(\vecn\pm\vece_1) p(\vecn\pm\vece_1,t)$ in \eq{dotp1nt}. In this approximation, we effectively neglect the backaction of the first channel  on all the other $N-1$ channels. 

In this mean-field type approach, the effect of the other channels is that of an external `clock' as expressed by the time-dependence of 
$
\langle n_{\rm av} \rangle_t = (\gamma_+ - \gamma_-) t \equiv I t
$
according to  \eq{1stmomentresult}. This leads to
\be\label{dotp1nta}
\dot{p}_1(n,t) &=&  \sum_{\pm} \left[ f_\mp\left( It - n\mp 1\right) p_1(n\pm 1,t)\right.\nonumber\\
 &-& \left. f_\mp\left( It - n\right) p_1(n,t)\right],
\ee
which describes non-autonomous feedback control as in \cite{Bra10} via rates \eq{gammasynch} with the `control current' $I=\gamma_+ - \gamma_-$.
As mentioned already after \eq{CLLMF}, the second cumulant \eq{secondcumnonauto} of the model \eq{dotp1nta} co--incides with  that of the constant interaction $N$-channel model, 
\eq{CLLMF}, for $N\to \infty$. In that limit and within the Gaussian approximation, the correspondence therefore is exact.

\section{Noise Spectrum}
The symmetrized noise spectrum $S_{ll'}(\omega)$, \eq{Snoise}, follows from the MacDonald formula as outlined, e.g.,  in \cite{LAB07}. One starts from
\be
S_{ll'}(\omega) = \int_{-\infty}^\infty d\tau e^{i\omega \tau} \langle  \delta I_l(t+\tau)  \delta I_{l'}(t) + (l\leftrightarrow l')   \rangle,
\ee
which still depends on the initial time $t$ that is sent to infinity at the end. We write
\be
\int_t^{t+\tau}dt' \delta I_l(t') = \delta n_l(t+\tau) - \delta n_l(t),
\ee
where $\delta n_l(t) \equiv n_l(t) - t I_l$ with $tI_l = \langle  n_l(t)\rangle$, cf. \eq{1stmomentresult}. As in \cite{LAB07}, this leads to 
\be\label{MacDonald}
& &\frac{S_{ll'}(\omega)}{2\omega} = \int_0^\infty d\tau \sin (\omega \tau) \\
&\times &
\frac{\partial}{\partial \tau}  \left\langle \left[ \delta n_l(t+\tau) - \delta n_l(t)\right]   \left[ \delta n_{l'}(t+\tau) - \delta n_{l'}(t)\right]  \right\rangle.\nonumber
\ee
To proceed, we now need (in the mixed terms in \eq{MacDonald}) the two-time correlation functions $\left\langle \delta n_l(t) \delta n_{l'}(t+\tau)\right\rangle$,
which in view of \eq{1stmoment} fulfill the regression equations
\be
& & \frac{d}{d\tau} \left\langle \delta n_l(t) \delta n_{l'}(t+\tau)\right\rangle  =\\
&=& \gamma \sum_{j=1}^N g_{l'j} \left( \left\langle \delta n_l(t) \delta n_{j}(t+\tau)\right\rangle - (j\to l') \right). \nonumber
\ee
We can solve this by first transforming into Fourier $(\veck)$ space similar to what we did to obtain the equal time correlation function $C_{ll'}(t)= \left\langle \delta n_l(t) \delta n_{l'}(t)\right\rangle  $ in \eq{Clj}, which in fact serves as the initial condition at $\tau=0$ here. The result 
\be
& &\left\langle \delta n_l(t) \delta n_{l'}(t+\tau)\right\rangle = \\
\frac{\gamma t}{N} &+&
\frac{\gamma}{N}\sum_{\veck\ne 0} \frac{e^{-i\veck(\vecR_l -\vecR_{l'})}}{\Gamma_\veck} \left( 1- e^{-\Gamma_\veck t}\right) e^{-\frac{\Gamma_\veck}{2} \tau} \nonumber
\ee
can now be used in the MacDonald formula \eq{MacDonald}, leading to 
\be
& &S_{ll'}(\omega) =2\omega \int_0^\infty d\tau \sin (\omega \tau) \\
&\times & \left[\frac{\gamma}{N} + \frac{\gamma}{N} \sum_{\veck\ne 0}  \cos \left({\veck(\vecR_l-\vecR_{l'}) }\right)  e^{-\frac{\Gamma_\veck}{2} \tau}      \right].
\ee
Here, as usual  \cite{LAB07} the sine function under the integral  in the first term has to be understood as  the limit $\sin \omega\tau = \lim_{\eta\to 0}\Im e^{i(\omega -\eta)\tau}$, and we now obtain \eq{noisefinal}.

\section{Information flow}
From \eq{infodef}, the temporal change of the mutual information $\mathcal{I}(t)$ is 
\be
\dot{\mathcal{I}}(t) =  -\sum_{l=1}^N \frac{\partial}{\partial x_l} J_l(\vecx,t) \left[ \ln P - \ln P_1 - \ln P_{1'}\right],
\ee
where we used \eq{FP} and the conservation of probabilities, $\int dx_1  \dot{P}_1(x_1,t)=0$, $\int dx_2...dx_N \dot{P}_{1'}(x_2,...,x_n,t)=0$, $ \int d^N\vecx \dot{P}(\vecx,t)=0$. We can write
\be
&&\int d^N\vecx \left(\frac{\partial}{\partial x_1} J_1(\vecx,t) +\sum_{l=2}^N \frac{\partial}{\partial x_l} J_l(\vecx,t) \right) \ln P_1(x_1,t) \nonumber\\
&=& \int d^N\vecx \left( \frac{\partial}{\partial x_1} J_1(\vecx,t) \right) \ln P_1(x_1,t)
\ee
by integration of parts of the second term, and similar for  $\ln P_{1'}$. This yields the splitting $\dot{\mathcal{I}}(t) \equiv \iota_{1'\to 1} +  \iota_{1\to 1'}$ with the information flows
\be
 \iota_{1'\to 1}&\equiv& - \int d^N\vecx \left( \frac{\partial}{\partial x_1} J_1(\vecx,t) \right)  \ln \frac{P}{P_1}\nonumber\\
 \iota_{1\to 1'}&\equiv& - \int d^N\vecx \left( \sum_{l=2}^N \frac{\partial}{\partial x_l} J_l(\vecx,t) \right) \ln \frac{P}{P_{1'}}.
\ee
Using the definitions of the global and local Shannon entropies, \eq{Shannon} and \eq{S1def}, we can now write the information flow $\iota_{1'\to 1}$ from channels $2,...,N$ to channel $1$ in the 
form \eq{info}.

For the evaluation of $\frac{d}{dt}S_1(t)$, \eq{S1def}, we use integration by parts to write
\be
\frac{d}{dt}S_1(t)&=& -\left\langle F_1(\vecx)  \frac{ \frac{\partial}{\partial x_1}  P_1   }{ P_1} \right\rangle
- \frac{\gamma}{2}\left\langle \frac{\partial}{\partial x_1}   \frac{ \frac{\partial}{\partial x_1}  P_1   }{ P_1} \right \rangle.
\ee
We use the fact that  $P_1(x_1,t)$ has to be a Gaussian,  and we thus obtain (with the definition of $F_1(\vecx)$ in \eq{FP}) the explicit form
\be
\frac{d}{dt}S_1(t) &=& \frac{1}{\langle \delta x_1^2 \rangle} \sum_{j=1}^N g_{1j} \left\langle   (\delta x_j-\delta x_1) \delta x_1 \right\rangle 
+ \frac{\gamma}{2\langle \delta x_1^2 \rangle}\nonumber\\
&=& \gamma \frac{ \sum_\veck e^{-\Gamma_\veck t}      }{2    \sum_\veck \sigma_\veck    }= \frac{1}{2}\frac{d}{dt} \ln \sum_\veck \sigma_\veck(t),
\ee
where $ \delta x_1 = x_1 -  (\gamma_+ - \gamma_-)t$ and we used Fourier components, $\langle \delta x_1^2 \rangle = \frac{1}{N} \sum_\veck \sigma_\veck$, and the second cumulants $\sigma_\veck(t)$ in $\veck$ space, \eq{Gammarates}. 

For the global Shannon entropy change, in a similar manner (by exploiting the equivalence of all channels $l$ in the homogeneous case) we obtain
\be
\frac{1}{N}\frac{d}{dt} S(t) &=& \int d^N\vecx   \frac{\partial}{\partial x_1} J_1(\vecx,t)        \ln P(\vecx,t)\\
&=&\frac{1}{2N}\sum_\veck  \frac{\Gamma_\veck e^{-\Gamma_\veck t}}{1-e^{-\Gamma_\veck t}} = \frac{1}{2N}\frac{d}{dt}  \sum_\veck \ln \sigma_\veck(t), \nonumber
\ee
where again we used the fact that the full distribution function $P$ is a Gaussian.
\end{appendix}


\begin{thebibliography}{10}

\bibitem{Zhangetal2015}
J. Zhang, Y. Liu, R. Wu, K. Jacobs, F. Nori, arXiv:1407.8536 [quant-ph] (2014). 

\bibitem{WM2009} H. M. Wiseman and G. J. Milburn, {\em Quantum Measurement and Control} (Cambridge University Press, Cambridge, England, 2009).

\bibitem{TSUMS2010}
S. Toyabe, T. Sagawa, M. Ueda, E. Muneyuki, and M. Sano, Nat. Phys. 6, 988 (2010).

\bibitem{SEKB11}
G. Schaller, C. Emary, G. Kie{\ss}lich, and T. Brandes, Phys. Rev. B 84, 085418 (2011).

\bibitem{AMP2011}
D. V. Averin, M. M\"ott\"onen, and J. P. Pekola, Phys. Rev. B 84, 245448 (2011).

\bibitem{AS2011}
D. Abreu and U. Seifert, Europhys. Lett. \textbf{94}, 10001 (2011).

\bibitem{JB2012}
Y. Jun and J. Bechhoefer, Phys. Rev. E 86, 061106 (2012).

\bibitem{BGK2013}
J. Bergli, Y. M. Galperin, and N. B. Kopnin, Phys. Rev. E \textbf{88}, 062139 (2013).

\bibitem{Kosetal2014}
J. V. Koskia, V. F. Maisia, J. P. Pekola, and D. V. Averin, PNAS  \textbf{111} (38),  13786 (2014).

\bibitem{LGK2014}
S. A. M. Loos, R. Gernert, and S. H. L. Klapp, Phys. Rev. E  \textbf{89}, 5 (2014).




\bibitem{SU2008}
T. Sagawa and M. Ueda, Phys. Rev. Lett. \textbf{100}, 080403 (2008).

\bibitem{SU2010}
T. Sagawa and M. Ueda, Phys. Rev. Lett. \textbf{104}, 090602 (2010).

\bibitem{HP2011}
J. Horowitz and J. M. P. Parrondo, Europhys. Lett. \textbf{95}, 10005 (2011).

\bibitem{SU2012}
T. Sagawa and M. Ueda, Phys. Rev. E \textbf{85}, 021104 (2012).

\bibitem{AS2012}
D. Abreu and U. Seifert, Phys. Rev. Lett. \textbf{108}, 030601 (2012).

\bibitem{MR2012}
T. Munakata and M. L. Rosinberg, J. Stat.Mech. (2012) P05010.

\bibitem{ES2012}
M. Esposito and G. Schaller, Europhys. Lett. \textbf{99}, 30003 (2012).

\bibitem{HSP2013}
J. M. Horowitz, T. Sagawa, and J. M. R. Parrondo, Phys. Rev. Lett. \textbf{111}, 010602 (2013).

\bibitem{SSEB2013} 
P. Strasberg, G. Schaller, T. Brandes, and M. Esposito, Phys. Rev. Lett. \textbf{110}, 040601 (2013).

\bibitem{Tas2013}
H. Tasaki, arXiv:1308.3776 (2013).

\bibitem{HE2014}
J. M. Horowitz, M. Esposito, Phys. Rev. X \textbf{4}, 031015 (2014)

\bibitem{HBS2014}
D. Hartich, A. C. Barato, U. Seifert, J. Stat. Mech. P02016 (2014). 

\bibitem{BS2014}
A. C. Barato and U. Seifert, Phys. Rev. Lett. \textbf{112}, 090601 (2014). 

\bibitem{SS2015}
N. Shiraishi and T. Sagawa, Phys. Rev. E \textbf{91}, 012130 (2015).


\bibitem{SSBJ2014}  P. Strasberg, G. Schaller, T. Brandes, and C. Jarzynski, Phys. Rev. E 90, 062107 (2014).

\bibitem{Hor2015}
J. M. Horowitz, arXiv:1501.05549 (2015). 

\bibitem{Gri2015}
A. L. Grimsmo, arXiv:1502.06959 (2015). 



\bibitem{Lloyd2000}
S. Lloyd, Phys. Rev. A \textbf{62}, 022108 (2000). 

\bibitem{Caretal2013}
A. Carmele, J. Kabuss, F. Schulze, S. Reitzenstein, and A. Knorr, Phys. Rev. Lett. \textbf{110}, 013601 (2013).

\bibitem{HSCK2014}
S. M. Hein, F. Schulze, A. Carmele, and A. Knorr, Phys. Rev. Lett. \textbf{113}, 027401 (2014).

\bibitem{GPS2014}
A. L. Grimsmo, A. S. Parkins, and B.-S. Skagerstam, New J. Phys. \textbf{16}, 065004 (2014).

\bibitem{KESB2015}
W. Kopylov, C. Emary, E. Sch\"oll, T. Brandes, New J. Phys. \textbf{17}, 013040 (2015).

\bibitem{EG2014}
C. Emary and J. Gough, Phys. Rev. B \textbf{90}, 205436 (2014).



\bibitem{Bra10} T. Brandes, Phys. Rev. Lett. 105, 060602 (2010).

\bibitem{Nazarov2013}
{\em Quantum Noise in Mesoscopic Physics}, edited by Y. V. Nazarov (Kluwer Academic, Dordrecht, 2003), Vol. 97.


\bibitem{IS2013} S. Ito, T. Sagawa,  Phys. Rev. Lett. 111, 180603 (2013). 



\bibitem{BruusFlensberg} H. Bruus and K. Flensberg, {\em Many-Body Quantum Theory in Condensed Matter Physics}, Oxford University Press (2004). 

\bibitem{HH77} P. C. Hohenberg and B. I. Halperin, Rev. Mod. Phys. \textbf{49}, 435 (1977). 

\bibitem{KardarBook} M. Kardar, {\em Statistical Physics of Fields}, Cambridge University Press (2007).  

\bibitem{MacDonald} D. K. C. MacDonald, Rep. Prog. Phys. 12, 56 (1948).

\bibitem{LAB07} N. Lambert, R. Aguado, and T. Brandes, Phys. Rev. B \textbf{75}, 045340 (2007).

\bibitem{Carmichael_02} H. L. Carmichael, {\em Statistical Methods in Quantum Optics 1}, Springer, Berlin (2002). 

\bibitem{Tom2006} T. Tom\'{e}, Braz. J. Phys. \textbf{36}, 1285 (2006). 

\bibitem{AJM2009} A. E. Allahverdyan, D. Janzing, G. Mahler, J. Stat. Mech. P09011 (2009).

\bibitem{AN2015} M. H. Ansari and Y. V. Nazarov, arXiv:1502.08020 (2015). 




\bibitem{ultracold}
S. Krinner, D. Stadler, D. Husmann, J.-P. Brantut, and T. Esslinger, Nature \textbf{517}, 64 (2015). 

\end{thebibliography}
\end{document}